\documentstyle[12pt,draft]{article}
 \newcommand \be {\begin{equation}}
\newcommand \bea {\begin{eqnarray} \nonumber }
\newcommand \ee {\end{equation}}
\newcommand \eea {\end{eqnarray}}

\newcommand{\bbi}{ \noindent}

\begin{document}

\title{Complexity in Biology:
The Point of View of a Physicist \\
{\small Talk given at {\sl Thinking Science for Theaching:
the Case of Physics}}\\
{\small Rome September 1994}
}
\author{  Giorgio Parisi \\
Dipartimento di Fisica,
Universit\`a {\sl La  Sapienza}\\ INFN Sezione di Roma I \\ Piazzale
Aldo Moro, Roma 00187}
\maketitle

\begin{abstract} We will review some of the theoretical progresses that
have been in the study of complex systems in physics and of their
applications to biology.
 \end{abstract}

\section{On the Definition of Complexity}

In recent years many progresses have been done in understanding the behaviour
of complex systems from the physical point of view (M\`ezard et al. 1987,
Parisi 1992) and now we can start to reflect on the possible applications of
these findings in a biological setting.

Many meanings can be attached to the word complex (Peliti and Vulpiani 1988).
In order to understand the precise meaning of the word complex that I use in
this talk it is useful to consider a system  composed by many parts at the
microscopic level. Very often we can describe it also at the macroscopic
level. The crucial point is the richness of this macroscopic description. Let
me present two examples: if our system is water, at the macroscopic level we
can only specify its temperature, pressure and normally there are no
remaining ambiguities; only for particular values of the temperature and
pressure there is the possibility of coexistence of more than one phase, e.g.
liquid and solid at zero centigrade.

In other words for a physical system the macroscopic description corresponds
to state in which phase the system is (e.g. solid, liquid and gaseous). At a
given temperature and pressure the number of choices is very small and
consequently the macroscopic description is very poor. On the contrary, if
our system is a biological system, its macroscopic description may be
extremely rich (an animal may run, sleep, eat, jump.....). The variety of the
macroscopic description will be taken as an indication of complexity. In
other words a system is complex if its macroscopic description is rich.

\section {The New Approach to Physical Complexity}

{}From this point of view clearly all traditional physical systems are simple
and apparently complexity is not relevant in the world of physics. In these
last years the situation has changed: it has been found that there are many
disordered physical systems for which the macroscopic description is quite
rich.

An example that is easy to visualise is an heteropolymer, i.e. a polymer
composed by a sequence of many different functional units. Typical biological
heteropolymers are proteins, DNA and RNA. Sometimes it happens that the same
heteropolymer at low temperature folds in a unique way, but in other cases
more than one folding is possible. If the heteropolymer may fold in many
different ways, we can consider each folding as a different phase and such a
system it is a complex system (Shaknovich  et al., Iori et al).  Other
physical systems, like spin glasses, have similar properties; they have been
carefully investigated and structure of different phases of the systems has
been studied in details. These findings on the existence of physical
structures with a complex macroscopic description open the way to the
construction of a physical theory of complexity.

It is striking  that most these systems have a rather interesting chaotic
behaviour: a small change in the form of the system may completely upset the
macroscopic behaviour. This phenomenon is well known in biology: a single
substitution of an aminoacid in a protein, may changing its folding
properties and upset its functionality. A single mutation is a living system
may have dramatic effects.

If a small change in the system deeply influences it macroscopic behaviour,
and this effect becomes more and more relevant by increasing the size of the
system, in the case of large systems the macroscopic behaviour is extremely
difficult to predict because it is sensitive to a huge number of microscopic
details. This phenomenon is very well known to all the people that have tried
to compute (in most cases unsuccessfully) from first principles the way in
which the protein folds.

This kind of difficulty is not without precedent in physics. Indeed the
observation that for a given system the actual trajectory is extremely
sensitive to the initial conditions (think of billiard balls), destroyed the
hope of computing the trajectory of a large system in a precise way (apart a
few exceptions). However the born of statistical mechanics is related to this
difficulty; the unpredictability of the trajectory in a deterministic sense
makes possible to obtain probabilistic predictions for the behaviour of the
system for generic initial conditions. The main proposal of Boltzmann was to
give up the possibility of predicting the evolution of the system for given
initial conditions and to concentrate the attention on the study of the
mostly likely evolution starting from generic initial conditions.

In the same way we can give up the possibility of computing which are all the
macroscopic descriptions of a particular complex system (M\`ezard et al. 1987,
Parisi 1992). Doing so we gain the possibility of obtaining statistical
predictions on it behaviour. The statistical predictions however are
different from the ones of usual statistical mechanics. In usual statistical
mechanics the system is nearly always in one given macrostate and we compute
the probability distribution of the several different microscopic
realisations of the same macrostate. Here we predict the probability of
having a given number of simultaneously available macrostates and the
relations among the different macrostates. Other interesting quantities can
be computed, for example the average time spent by a system in a given
macrostate before jumping in an other macrostate.

It is remarkable that the for a large class of systems, a generic choice of
the system implies the existence of many macroscopic states. In other words
if the system is chosen in a random way, the macroscopic behaviour is
automatically very rich. We do not need to tune the parameters that control
our system in order to have many different macroscopic states because this
feature is present in the generic case. We can summarise the situation by
saying that microscopic randomness generates complexity.

\section {The Role of Chance in Biology}

This approach may be very useful in the case of biology where very often we
have systems that have a very rich macroscopic behaviour that cannot be
easily explained in terms of the microscopic composition: for example
$10^{12}$ atoms give rise a living cell, $10^{10}$ cells give rise to a brain
and I do not know how many living beings give rise to an ecosystem.

This approach may be useful in biology as far as the role of chance is
important in determining the present form of a living object.

Let me give an example on the role of chance outside biology. We consider a
computer. We can schematise a computer as a set of electronic components
(transistors, diodes, resistors, capacitors...) connected one with the others
and with an input/output channel. We can do the theory of the behaviour of a
hypothetical set of randomly connected elements (Carnevali et al 1987),
however this theory is completely useless and irrelevant for the
understanding of the properties of real computers. Indeed computers are not
constructed by assembling in random way electronic components, but they have
been designed in such a way to work. In most of the case any random
displacement from the original design leads to a disaster.

One could suppose that a computer has been designed by choosing randomly one
among of the possible computers that work in the planned manner. Also this
refined conjecture is not true: real computers are designed by men and
constructed in such a way that men can understand how they work.

The same problem exists in biology. Have the living beings been randomly
chosen among all possible forms that do survive or has the natural evolution
selected the organisms according to other principles? If the assumption that
living beings been randomly chosen among all the possible ones is not a bad
approximation to reality, the previous considerations may be useful in
biology, otherwise they would be irrelevant in most of the problems of this
field.

Much depends upon the underlying structure of the living organism, although a
major disagreement exists over this point among different approaches. The
cell is often seen as a larger computer with DNA representing the program
(software) and the proteins representing the electrical circuits (hardware).
If this idea was not wide of the mark there would be no point in using
statistical mechanics to study biology, just as there is no point in using it
to study a real computer. A living organism is not made in a totally random
fashion but equally it is not designed on paper. Living organisms have evolved
via a process of random mutation and selection.

These two aspects are crucial to the study of protein dynamics. On the one
hand it is clear that proteins have a well defined purpose and have been
designed to achieve it. However, proteins have initially been generated in a
random manner and perhaps some of the physical properties of proteins
(particularly those which have not been selected against) still reflect the
properties of polypeptide chain with elements chosen at random along the
chain.

The marriage of determinism and chance can be found if we study the
development of a single individual. For example, the brains of two twins may
appear completely identical if not examined under the microscope. However,
the positions and the connections of the neurons are completely different.
Individual neurons are created in one part of the cranium, migrate to their
final position and send out nerve fibres that attach themselves to the first
target they reach. Without specific signals on the individual cells such a
process is extremely sensitive and therefore the slightest disturbance leads
to systems with completely different results for the individual connections.
The metaphor of the computer does not seem adequate in that the description
of the fine detail (the arrangement and the connection of the individual
elements) is not laid down in the initial design. Moreover, the number of
bits of information required to code the connections in a mammal brain is of
the order of $10^{15}$, far grater that the $10^{19}$ bits of information
contained within DNA.

The arrangement of the neurons and of their connection in the brain during
the ontogenesis is an excellent example of a disordered  system, in which
there is a deterministic, generically controlled components (all that is the
same in the brains of two twins, i.e. the external form, weight, possible the
hormonal balance) and a chance element which differ from twin to twin, Our
attitude toward the methodology that should be used to achieve an
understanding of the behaviour of the brain changes completely upon whether
we consider the variable (and therefore chance) part to be a non essential,
non functional accident or if we thing that some characteristic of the
variable part are crucial for proper functions.

\section{Memory and Learning}

The use of techniques for the statistical mechanics of disordered systems is
certainly useful in the case of learning and more generally in order to
modelling the reaction of an organism with the outside word.  General
speaking the information that arrive from the outside word a both a random
and deterministic component. For example the face of the people we know have
a constant component (they are faces, not zucchini) and a variable (chance)
one, the characteristic of each individual.

The input signal produces modifications of the brain (mostly variation of the
synaptic strengths) and the chance nature of the events being memorised
implies the random nature of  the growth of synapses between the various
neurons and therefore a disordered synaptic systems. The goal of many
investigations it to understand how some of the modifications of the brain
are useful for memorise the past experience. Moreover the brain can also
examine its experience for finding regularities and constructing models that
enable it to predict the future. Of course this possibly of predicting (at
least partially) the future implies that our experience is not fully random,
but it satisfies certain rules.

Generally speaking the previously described approach can be used to
understand how systems may behaviour in presence of inputs that have both a
regular and a random component. The most interesting case is when one studies
theoretically systems which are constructed in such a way to have a
resemblance with the behaviour of real neurons. One can assemble these formal
neurons in such a way that they learn to remember the past experience, to
organise it in categories and sometimes to guess some of rules that are
satisfied by the inputs. These neural networks have been extensively studied
(Amit 1987), especially after the seminal paper of Hopfield (Hopfield 1982),
and a special attention has been paid to the way in which they can work as a
memory. This particular issue is quite well understood, at least in some
model systems.

The most interesting and open problem is how to construct networks that can
generalise a rule from examples (Carnevali et al. 1987, Denker et al. 1987,
Seung et al. 1991). Here things become more interesting and subtle: a given
set of examples may be not enough to determine the rule; the rule may have
exceptions, which are learned at a later stage than the rule itself. It also
possible to compare how different neural networks generalise a rule, which
are the mistakes they make and how the steps of this kind of learning are
related to the steps that human beings do when confronted to the same
problem. This field is rapidly developing. Its pratical applications (e.g. in
pattern recognition) are extremely promising, moreover it is likely that it
will play a more and more important role in our understanding of the way we
think and learn.

\section{References}

 D.J. Amit  (1989), {\sl Modeling Brain Functions}, Cambridge
University  Press, Cambridge.

\bbi Denker J., Schwartz D., Wittner B., Solla S., Horward S., Jackel L. and
Hopfield J.J. (1987), {\sl Automatic learning}, Complex Systems, {\bf 1
}, 877-888.

\bbi Carnevali P. and Patarnello S. (1987) {\sl Boolean Networks which Learn to
Compute}, Europhys. Lett., {\bf  4
}, 1199-1204.

\bbi Hopfield J.J. (1982) {\sl Neural networks and physical systems with
emergent  collective computational abilities,} Proc. Natl. Acad. Sci. USA  {\bf
79},  2554-2558.

\bbi G. Iori, E. Marinari and G. Parisi,  {\sl Random Self-Interacting Chains:
a
Mechanism for Protein Folding}, J. Phys. A (Math. Gen.)  {\bf 24} (1991)
5349.

\bbi M. M\`ezard, G. Parisi and M.A. Virasoro (1987),  {\sl Spin glass theory
and  beyond}, World Scientific.

\bbi Parisi G. (1992)  {\sl Order, Disorder and Simulations}, World Scientific,
Singapore.

\bbi Peliti L. and Vulpiani A. Eds. (1988)  {\sl Measures of Complexity},
(Springer-Verlag Berlin).

\bbi H.S. Seung, H. Sompolinsky and N. Tishby,  {\sl Statistical Mechanics of
Learning from Examples}, Phys. Rev. A.  {\bf 33}, 1978-1982 (1991).

\bbi Shaknovich E.I. and Gutin A.M. (1989)  {\sl The Nonergodic (Spin Glass
Like)  Phase of Heteropolymer with Quenched Disordered Sequence of Links},
Europhys.  Lett.  {\bf 8}, 327-332.
\end{document}